\documentclass[aps,prl,twocolumn]{revtex4}
\usepackage{amsmath}
\usepackage{amssymb}
\usepackage{graphicx}

\begin{document}

\title{Optical limiting using Laguerre-Gaussian beams}

\author{Weiya Zhang}
\email{weiya_zhang@wsu.edu}

\author{Mark G. Kuzyk }
\email{kuz@wsu.edu}

\affiliation{Department of Physics and Astronomy, Washington State University, Pullman, WA 99164-2814}

\date{\today}

\begin{abstract}
We demonstrate optical limiting using the self-lensing effect of a higher-order Laguerre-Gaussian beam in a thin dye-doped polymer sample, which we find is consistent with our model using Gaussian decomposition. The peak phase shift in the sample required for limiting is smaller than for a fundamental Gaussian beam with the added flexibility that the nonlinear medium can be placed either in front of or behind the beam focus.
\end{abstract}

\maketitle

\noindent With the availability of ever more intense light sources, optical limiters -- which can protect sensors and eyes from optical damage -- are becoming of great interest.\cite{Tutt:1993,Lepkowicz:2002,He:1995, Ehrlich:1997,Jia:2004,Leite:1967, Pan:2006,Gary:1999} Passive optical limiters are often preferred over active ones due to their simplicity and fast response. Many mechanisms have been proposed to make passive optical limiters, including nonlinear absorption such as reversed saturable absorption,\cite{Lepkowicz:2002} two photon absorption,\cite{He:1995, Ehrlich:1997} free carrier absorption,\cite{Jia:2004} nonlinear refraction,\cite{Leite:1967} induced scattering,\cite{Jia:2004, Pan:2006} and photorefraction.\cite{Gary:1999}

One promising optical limiter design is based on self lensing, which originates in the nonlinear refractive index of a material.\cite{Leite:1967,Tutt:1993} The refractive index changes in proportion to the intensity of the incident light and forms a spacial distribution according to the beam profile of the incident beam.
If the incident beam is a fundamental Gaussian beam, the optical `thickness' of a thin film of such a medium follows the intensity profile and thus forms a positive or a negative lens depending on the sign of the refractive index change.
\begin{figure}[htb]
\centerline{\includegraphics{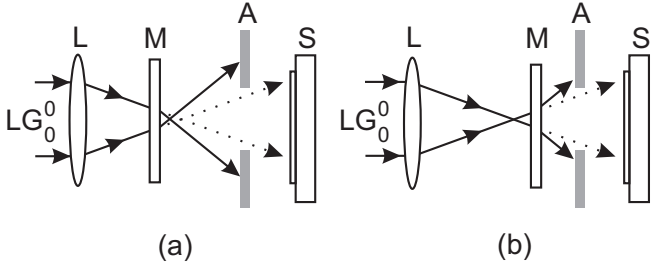}}
\caption{ \label{fig:selflensingofLG00} Schematic diagram of a self-lensing limiter, where the incident beam is assumed to be a fundamental Gaussian beam ($LG_0^0$). (a) Self focusing. (b) Self defocusing. L: lens, M: nonlinear medium,  A: aperture, and S: sensor.}
\end{figure}

Figure~\ref{fig:selflensingofLG00} shows an optical limiter based on thin-film self lensing.  The nonlinear refractive medium is placed near the focus of the incident fundamental Gaussian beam, which goes through a small aperture in the far field, where it is detected with an optical sensor. At low enough beam intensity, nonlinear refraction is negligible, so most of the beam's energy passes the aperture and reaches the sensor. If the beam intensity is increased, self lensing increases beam divergence beyond the focal point.  As a result, a large portion of the energy is blocked by the aperture, thus protecting the sensor. Note that the location of the nonlinear medium is critical in order to make optical limiting effective.\cite{Tutt:1993} The nonlinear medium must be placed in front of the beam focus for the case of self focusing, and behind the beam focus for self defocusing.

While conventional optical limiters based on self lensing almost always use fundamental Gaussian beams (or similar ones, such as flat-top beams\cite{Kovsh:1999}), we report in this paper an optical limiter using a higher-order Laguerre-Gaussian (LG) beam. One of the advantages of higher-order LG beams is that the nonlinear medium can be placed either in front of or behind the focal plane without regard to the sign of the change of the refractive index; and as we show, the required peak phase shift in the material needed for limiting is lowered.

LG beams are a complete set of solutions of the wave equation in cylindrical coordinates.\cite{Kogelnik:1966} Each member of the set is characterized by two mode numbers, the angular mode number, l (l = 0, $\pm$1, $\pm$2, ...), and the transverse radial
mode number, p (p = 0, 1, 2, ...), written as $LG_p^l$. The lowest order, $LG_0^0$, is the fundamental Gaussian beam. When $l\geq1$, the LG beam possesses well defined orbital angular momentum of $l\hbar$ per photon,\cite{Allen:1992} helical phase fronts, and zero  light intensity and phase singularity on the beam axis. Higher-order LG beams have been recently applied to optical manipulation of microscopic particles,\cite{Paterson:2001} quantum information processing,\cite{{Mair:2001},{Barreiro:2003}} orbital angular momentum sorting,\cite{Wei:2003} spiral imaging,\cite{Torner:2005} scattering,\cite{Schwartz:2005} interference,\cite{Sztul:2006} and Z-scan measurements.\cite{Zhang:2006}  In this work, we show how a self-lensing limiter applies to higher-order LG beams and that such a limiter has advantages over one that operates on fundamental Gaussian beams.  Any beam profile can be converted to higher-order LG beams by, for example, dividing it with an array of apertures into pixels followed by conversion of each element with spiral phase plates. So, devices can be designed that are more generally applicable to arbitrary beams and images.

We focus on the $LG_0^1$ beam. The geometry is the same as in Figure~\ref{fig:selflensingofLG00} except the incident beam is a $LG_0^1$ beam. Ray optics adds no insights because of the complexity of the induced lens.  We thus calculate the far-field on-axis normalized transmittance,
\begin{align}
T\left(Z ,\Delta \Phi \right) = \frac{{\left| {E \left(r \to 0,\phi, z \to \infty \right)} \right|^2 }}{{\left| {E \left(r \to 0,\phi, z \to \infty \right)|_{ \Delta \Phi \to  0}} \right|^2 }}, \label{Eq:normalized_T}
\end{align}
where we use cylindrical coordinates and assume that the beam axis is along the z direction with the beam focus at $z=0$,  $E \left(r ,\phi, z  \right)$ is the beam's electric field, $Z=z_s/z_r$ is the normalized medium position with $z_s$ being the position of the nonlinear medium along the beam axis and $z_r$ the Rayleigh length of the incident beam, and $\Delta \Phi$ is the nonlinear phase shift experienced by the beam when traversing the nonlinear medium. $\Delta \Phi$ is proportional to the change of the refractive index, $\Delta n$, which in turn is related to the intensity, $I$, of the light. $\Delta \Phi$ is a function of position in the material and follows the transverse intensity profile of the beam. Since the aperture is placed at the far field and only allows the near-axis light to pass, $T$ is approximately the aperture's transmittance. $T$ is normalized so that it is unity when the light intensity is weak, i.e. when $\Delta \Phi \to 0$. In optical limiting, $T$ decreases when the incident intensity increases.

For a Kerr medium, where $\Delta n = n_2 I$ ($n_2$ is the nonlinear refractive index), the generalized Gaussian decomposition method\cite{Zhang:2006} yields
\begin{align}
T\left(Z ,\Delta \Phi_{peak} \right) &= 1+ \frac{8 e \cdot Z\left(27+10Z^2-Z^4\right)}{\left(9+Z^2\right)^3}\Delta \Phi_{peak},\label{Eq:Toflg01}
\end{align}
where $\Delta \Phi_{peak}$ is the peak nonlinear phase shift in the medium when placed at position Z. The intensity is highest at the point in the material where $\Delta \Phi = \Delta \Phi_{peak}$. Optical limiting relies on $\left| \Delta \Phi_{peak} \right|$ being large, which is a challenge for applications that require low limiting transmittance therefore an intensity in the material that may be high enough to damage it.  An optical limiter is thus most effective when $\Delta \Phi_{peak}$ corresponds to a peak intensity that is just below the damage threshold of the material. The effectiveness of an optical limiter should thus be judged by the quantity $\left. \Delta T \right |_{\Delta \Phi_{peak} = \Delta \Phi_{damage}}$, that is, the change in transmittance should be as large as possible when the peak phase change in the material corresponds to an intensity near the damage threshold.

\begin{figure}[htb]
\centerline{\includegraphics{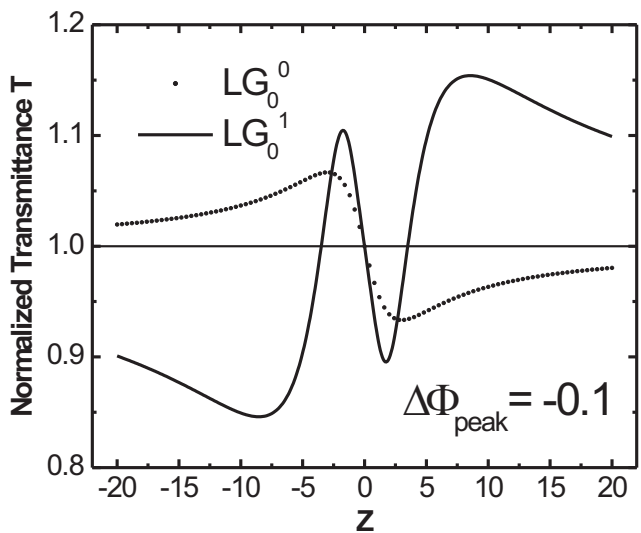}}
\caption{ \label{fig:Tlg0001} Far-field on-axis normalized transmittance, $T$, as a function of the the position of the nonlinear material, $Z$, of a self-defocusing medium for a a fixed value of $\Delta \Phi_{peak}=-0.1$. Solid and dotted curves correspond to a $LG_0^1$ beam and a $LG_0^0$ beam, respectively.}
\end{figure}
Figure~\ref{fig:Tlg0001} shows a typical theoretical plot of $T$ against the medium position $Z$ for a $LG_0^1$ beam and a $LG_0^0$ beam, where we assume a negative Kerr medium and $\Delta \Phi_{peak}=-0.1$ is fixed for all $Z$. For a self-focusing medium, the curves in Figure~\ref{fig:Tlg0001} for $\Delta \Phi_{peak}=+0.1$ would be the mirror image about Z=0. We emphasize that Figure~\ref{fig:Tlg0001} is not a Z-scan curve,\cite{Zhang:2006} nor is it a curve that can be easily determined experimentally because of difficulties associated with keeping $\Delta \Phi_{peak}$ constant (the incident beam intensity would need to be adjusted to do so).  This plot, however, is useful for determining the ideal position of the Kerr medium for optical limiting. In the curve representing the $LG_0^0$ beam, $T<1$ when $Z>0$, suggesting that the medium must be placed beyond the beam focus to get optical limiting, consistent with the ray diagram analysis in Figure~\ref{fig:selflensingofLG00} (b).  $Z = 3$ is the position of the largest decrease in transmittance, and therefore the optimum location for a limiter using a $LG_0^0$ beam.  For the $LG_0^1$ beam, there are two regions where T falls below 1. One is for $Z < \sim  -3.49$, the other is $ 0<Z< \sim 3.49 $. The optimum placement of the medium of an $LG_0^1$ beam is therefore around $Z=1.73$ and $Z=-8.55$.  Both of these positions yield more effective limiting because $\Delta T = 1-T$ is larger than for the fundamental Gaussian beam.  Furthermore, since optical limiting can take place when the nonlinear medium is placed either in front of or behind the beam focus, this provides more design flexibility.  For example, the limiter can be made more compact when the beam focus is behind the Kerr medium.

\begin{figure}[htb]
\centerline{\includegraphics{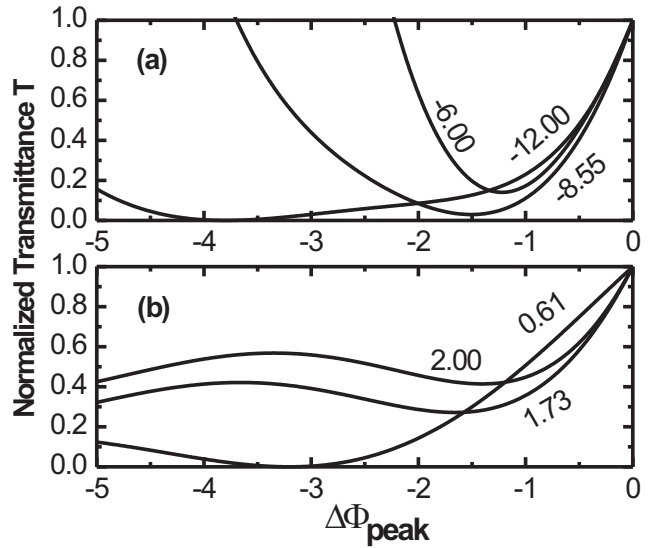}}
\caption{ \label{fig:LargePhi} The normalized transmittance, $T$, versus the peak nonlinear phase shift in the medium, $\Delta \Phi_{peak}$, for an $LG_0^1$ beam. The position of the medium, $Z$, for each of the curves is indicated by the number along that curve. (a) Medium in front of the beam focus. (b) Medium behind the beam focus.}
\end{figure}

The succinct form of Eq.~(\ref{Eq:Toflg01}) is obtained under the condition of $\left|\Delta \Phi_{peak}\right| << 1$, which is not necessarily true in optical limiting applications.  To exam the behavior of $T$ when $\left|\Delta \Phi_{peak}\right|$ is larger, we use numerical calculations. Figure~\ref{fig:LargePhi} shows some representative results, where $T$ is plotted against $\Delta \Phi_{peak}$. The number along each curve indicates the position, Z, of the medium associated with that curve.

For a wide range of $\Delta \Phi_{peak}$ values, $T$ decreases as $|\Delta \Phi_{peak}|$ increases. Thus, many sample positions can lead to efficient optical limiting when $\left|\Delta \Phi_{peak}\right| >> 1$. When the slope is steepest, the transmittance is the most strongly decreasing function of the input power, thus leading to more effective limiting and lower threshold power.  After $T$ reaches the minimum value, it turns back and increases as $|\Delta \Phi_{peak}|$ is further increased. This turnaround of $T$ is undesirable for optical limiting, therefore it sets an upper bound on the beam intensity, above which the transmittance again increases and limiting is lost. In practice, one should choose among these curves, i.e., choose an optimum position for the specific system requirements, such as limiting threshold, range of anticipated beam intensities, system geometry, etc.

We test our theory with experiments. We generate the $LG_0^1$ beam  by converting a linearly polarized $LG_0^0$ beam from a krypton laser (647 nm) using a computer-generated hologram (CGH).\cite{Zhang:2006} The nonlinear medium is a 1.4-mm-thick 2$\%$ w/w DR1/PMMA sample, whose nonlinearity is due to photoinduced trans-cis-trans isomerization of DR1 molecules followed by reorientation in the direction perpendicular to the polarization of the incident laser beam.\cite{Zhang:2002} Under low beam intensity and short exposure time, a DR1/PMMA sample can be approximately treated as an optical Kerr medium. When the beam intensity is high or exposure time is long, the change of refractive index in DR1/PMMA saturates.

\begin{figure}[htb]
\centerline{\includegraphics{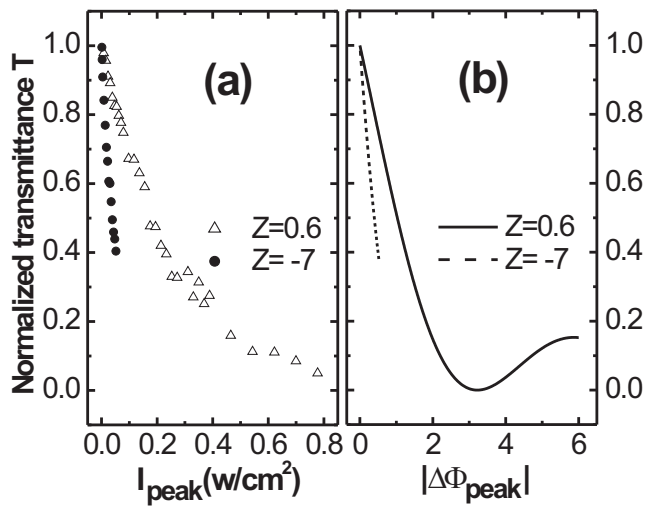}}
\caption{ \label{fig:z06zn7} Optical limiting using an $LG_0^1$ beam with the medium at $Z=0.6$ and $Z= -7$. (a) Experimental results with a DR1/PMMA sample. (b) Theoretical results assuming an optical Kerr medium.}
\end{figure}
Figure~\ref{fig:z06zn7}(a) shows the experimental results, where the normalized transmittance is measured after the DR1/PMMA sample is exposed for 300 seconds. $I_{peak}$ is the peak intensity in the sample when it is located at position $Z$. The intensity is kept below $I_{peak} \approx 50$ mw/cm$^2$ in the case of $Z=-7$ because the computer-generated hologram gets damaged at higher intensities.  Figure~\ref{fig:z06zn7}(b) shows the corresponding theoretical results based on an ideal optical Kerr medium model, where $\Delta \Phi_{peak}$ is calculated using the equivalent $n_2$ of the DR1/PMMA sample before saturation takes place. In general the experimental data follows the prediction of the theory, particularly for lower intensities. At higher intensities, the experimental data lags the theoretical transmittance, $T$, which one would expect is related to the saturation of the Kerr Effect in DR1/PMMA. We note that this lag has made optical limiting more effective at higher intensities than predicted by theory, which ignores saturation.

In conclusion, we have demonstrated optical limiting using self lensing of a $LG_0^1$ beam in a thin dye-doped polymer sample. The nonlinear medium can be placed either in front of or behind the beam focus without regards to the sign of the Kerr effect. At some sample positions, the peak phase shift in the sample required for limiting is smaller than for a fundamental Gaussian beam. Finally, although we have used the $LG_0^1$ beam as an example, we believe that optical limiting can also be observed in higher order $LG$ beams.  Since other beam profiles can be converted to an LG beam, it is possible to design limiters that operate on any type of beam.  Our results therefore add more choice and flexibility in system design that yield optical limiters with lower threshold intensity.

We thank the National Science Foundation (ECS-0354736) and Wright Paterson Air Force Base for generously supporting this work.

\end{document}